\documentclass[journal]{IEEEtran}

\ifCLASSINFOpdf
\else
   \usepackage[dvips]{graphicx}
\fi
\usepackage{url}
\usepackage{graphicx}
\usepackage{booktabs}
\usepackage{hyperref}
\usepackage{balance}

\begin{document}

\title{
Attention is All You Need? \\ Good Embeddings with Statistics are Enough:\\
Large Scale Audio Understanding without Transformers/ Convolutions/ BERTs/ Mixers/ Attention/ RNNs or ....}

\author{Prateek Verma
\thanks{This work has been submitted to the IEEE for possible publication. Copyright may be transferred without notice, after which this version may no longer be accessible; Submitted for review on 22 Jan 2022.}
\thanks{Prateek Verma is a Research Assistant with Stanford University, 450 Jane Stanford Way, Stanford CA 94305 (email: prateekv@stanford.edu).}}

\markboth{IEEE Signal Processing Letters, Vol. X, No. X, Month 20XX; }
{Shell \MakeLowercase{\textit{et al.}}: Bare Demo of IEEEtran.cls for IEEE Journals}

\maketitle
\begin{abstract}
This paper presents a way of doing large-scale audio understanding without traditional state-of-the-art neural architectures. Ever since the introduction of deep learning for understanding audio signals in the past decade, convolutional architectures have been able to achieve state-of-the-art results surpassing traditional hand-crafted features. In the recent past, there has been a similar shift away from traditional convolutional and recurrent neural networks towards purely end-to-end Transformer architectures. We, in this work, explore an approach, based on the Bag-of-Words model. Our approach does not have any convolutions, recurrence, attention, transformers, or other approaches such as BERT. We showcase an approach going against the mainstream research at the moment. We utilize micro and macro-level clustered vanilla embeddings and use an MLP head for classification. We only use feed-forward encoder-decoder models to get the bottlenecks of spectral envelops, spectral patches, and slices as well as multi-resolution spectra. A classification head (a feed-forward layer), similar to the approach in SimCLR is trained on a learned representation. Using simple codes learned on them, we show how we surpass traditional convolutional neural network architectures, and come strikingly close to outperforming powerful Transformer architectures. Our approach could have been carried out back in 2006, using the vanilla autoencoder and dimension reduction architectures, as proposed by Hinton et. al. \cite{hinton2006reducing}. We show that just by using this architecture, and doing simple statistics on the latent representations, we could have outperformed state of the art architectures audio understanding, like convolutional and recurrent architecture as late as 2018 \cite{fonseca2020fsd50k}. This goal of this work is, hopefully to would pave way for exciting advancements in the field of representation learning without massive, end-to-end neural architectures.
\end{abstract}


\IEEEpeerreviewmaketitle

%%%%% INTRO %%%%%%%%%%%%%%%%%%%%%%%%
\section{Introduction}
\label{sec:introduction}

Audio understanding is a widely studied problem in domains such as signal processing, machine learning, and perception. It is to teach computers to hear as humans do, and the goal is to help develop a human-level perception in machines. With the advancement in machine learning, there has been rapid progress in making these systems reach almost human-level performance. For this paper, we would pose the problem as that of understanding the categories of sound present in an input signal. CNN architectures have become a standard way of learning these mappings, as shown in \cite{hershey2017cnn, xu2018large,bian2019audio}. Further, with the recent success of Transformer architectures in computer vision \cite{dosovitskiy2020image,parmar2018image,he2021masked}, NLP \cite{vaswani2017attention,dai2019transformer}, \cite{wei2021finetuned} and audio \cite{verma2021audio,verma2021generative,dhariwal2020jukebox}, there has been a pivot recently on improvements on the core architecture, and adapting it for acoustic scene understanding \cite{gong2021ast}. There have been in the past approaches inspired by natural language processing such as computing a bag-of-words model \cite{zhang2010understanding} on features such as MFCC \cite{pancoast2012bag} yielding surprising strong results. We in this work take inspiration from the traditional bag-of-words model along with the powerfulness of neural net models, taking in some ways the best of two worlds. The current work is also based on the idea of learning codebook and doing some statistics on these learned code words as shown in \cite{verma2020framework}, \cite{baevski2019vq}, \cite{baevski2020wav2vec}. In problems such as unsupervised learning/one-shot learning \cite{wang2021calls}, the goal is to learn a latent representation for an input signal and use this latent representation for a variety of applications using a classification head \cite{chen2020simple,wang2021multi}, \cite{he2021masked}. Audio embeddings have been powerful to aid in a variety of applications such as speech recognition and audio understanding \cite{verma2019neuralogram,Chung2018-Speech2Vec}, \cite{haque2019audio}, conditional audio synthesis \cite{haque2019audio,skerry2018towards} as well as transformation \cite{oord2017neural}, \cite{verma2018neural}. This work also devises a way of coming up with a latent vector summarizing the contents of the audio signal and then using a classification head similar to \cite{chen2020simple, wang2021multi} for classification purposes. This work also draws inspiration from the idea of learning a mapping from a signal of interest to a clustered latent space as shown in \cite{verma2020unsupervised, aytar2016soundnet}, \cite{owens2016ambient}. Unlike their work which utilized UMAP \cite{mcinnes2018umap} or vanilla latent representations, we use coded book statistics as a representation of input mel-spectrogram. There have been approaches over discrete code words for representation learning such as \cite{sun2019videobert}, which trained a BERT over clustered image representations. 
\begin{figure*}[!hbtp] \centering \includegraphics[width=17.9 cm, height = 7 cm]{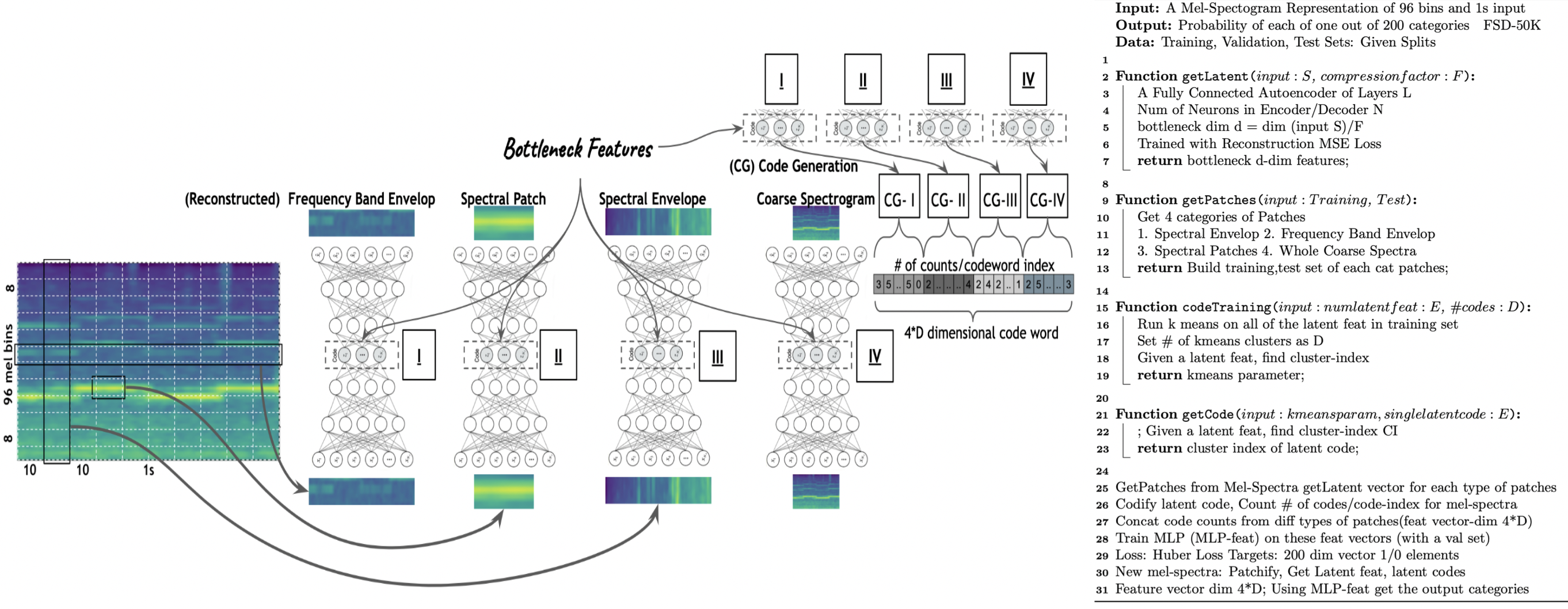} \caption{Figure capturing our proposed method and learning different codes for different representations of mel-spectogram. We divide up the audio spectrogram into 4 types of patches, namely frequency band envelop, spectral patches, spectral envelop, as well as overall coarse downsampled spectra. The manner in which we patch-up the spectrogram is as shown. We extract bottleneck latent representation for each of the type of patches, by training vanilla autoencoders. These features are individually clustered into $D$ clusters using k-means algorithm. Finally to extract a feature vector for the input spectrogram, we use count statistics for each type of cluter present for the patches. We concatenate each of the four code word statistics to yield the final feature vector. The total sum of the count in the $4*D$ feture vector is equal to the total number of different types of patches present (similar to how a bag-of-words model has the total number of words over dictionary elements as the sum in the feature vector \cite{bow}). On these $4*D$ dimensional vectors a typical train/val/test experiment is carried out and the results are reported. In order to be fair with comparison no data augmentation was carried out, and the setup was similar to \cite{fonseca2020fsd50k,verma2021audio}.} \label{fig:contour-pair} \end{figure*}

The contributions of the paper are as follows: i) We propose a framework purely based on learned embedding and statistics based on it and achieve significant performance as compared to traditional convolutional and transformer architectures. We achieve this without having any convolutional, transformer, mixer, attention, or recurrent blocks. ii) We compute statistics  (bag-of-words \cite{pancoast2012bag}) over dictionary learned over various latent representations of mel-spectrogram from vanilla autoencoders. We capture various facets of audio signals, and learn a concatenated dictionary of spectral patches, spectral envelop, frequency band envelops, and overall statistics. iii) We show how we can improve the performance of our model by randomly masking the input signal, making them robust by drawing inspiration from approaches such as BERT \cite{devlin2018bert}.

\section{Dataset}
\label{sec:dataset} For evaluation of our approach, we use Free-Sound 50K dataset \cite{fonseca2020fsd50k} instead of widely used AudioSet \cite{gemmeke2017audio}. It is an open dataset having over 100 hours of manually labeled samples drawn from the AudioSet ontology of 200 classes. This was done, as the audio samples are available as opposed to YouTube links, and the readers are advised to read \cite{fonseca2020fsd50k} for advantages of it over AudioSet \cite{gemmeke2017audio}. For the sake of this paper, we choose to report the results only on \cite{fonseca2020fsd50k} and expect similar results to hold true on other datasets. This is shown in \cite{gemmeke2017audio} how neural architectures having similar gains on AudioSet and FSD50K. One of the significant advantages is that it contains twice the same number of training examples in the balanced setup as AudioSet. We used the training, validation, and test split as given. The dataset has in total of about 51,197 clips. We have down-sampled all the clips to be of 16kHz. We may also mention that we have been consistent in the training and the validation setup while reporting the mean-average precision (mAP) scores as reported in \cite{fonseca2020fsd50k}. Training with a longer context will result in improved performance as well as augmentation etc., and the goal is to have a fair comparison of architectures in similar setup. The training was done on 1s of audio to predict one/mutiple categories of sound out of 200, with the label of the clip assigned to all of the 1s patches in case the clip is longer. Once the model is trained, the mAP scores are reported at clip level, with the probability scores averaged on 1s chunks to predict the contents (possible classes out of 200). Thus we can compare our performance with the baseline convolutional (DenseNet, Res-Net, VGGish) \cite{fonseca2020fsd50k} and Transformer architectures \cite{verma2021audio}. 

\section{Methodology}
This section describes the methodology of how we obtained a feature vector from a mel-spectrogram input. Briefly, for every input representation, the system learns a latent representation by fully-connected vanilla auto-encoders with varying degrees of compression factor $F$. Once we obtain these latent features, we cluster them according to a fixed number of dictionary elements having the vocabulary $D$ (equal to the number of clusters from k-means), assigning each input latent representation a discrete value from $0$ to $D-1$, according to the cluster to which it belongs. We obtain overall statistics of the counts of each of the dictionary elements similar to a bag-of-words (BOW) model. Multiple representations are concatenated, to obtain a final feature code of size $4*D$, which is used to train a MLP head (instead of spectrograms) in a typical training/val/test setup (similar to \cite{chen2020simple}). 
\subsection{Learning Latent Representations}
We choose to work with mel-spectrogram input having a total of 96 bins encompassing a total range from 0 to 8kHz, as our input signals are 16 kHz. We choose a hop size of 10ms, with a window of 30ms with FFT size to be 2048, with a hanning window.
For learning a latent representation for a given input, we deploy a three-layer MLP encoder with a bottleneck layer of size to be reduced by a factor of $F$ from the input dimension. For this paper, we experiment with compression factor $F = 10$ and $F = 20$. This bottleneck is then again passed onto a 3 layer MLP decoder to reconstruct back the same input. We used mean-square error criteria for training this auto-encoder block, with 2048 neurons of encoder and decoder with a dropout factor of 0.5. We call different encoder representations learned at a particular compression factor $F$, to be $e_F^{pat}$, $e_F^{env}$, $e_F^{fenv}$, $e_F^{o}$ for an encoding learned for a spectral patch, spectral envelop, frequency band energy across time, and scaled down-sampled mel-spectra.

\subsubsection{Representation of Spectral Patches}
To extract embeddings from spectral patches, we draw inspiration from recent works in audio/vision Transformer \cite{verma2021audio,dosovitskiy2020image}. We take patches of size 8 frequency bins along the frequency axis, with 10 bins along time, across the mel-spectrogram which has input dimensions of 96 x 100 for each of the 1s inputs. We divide the input spectrogram equally as shown in Fig.1. For all of the training data, auto-encoders are trained on these patches, to obtain a patch level embedding $e_F^{pat}$. This captures the local variations present in an input mel-spectrogram. By dividing the 1s spectrogram in a grid form, we obtain 120 patch level embeddings (96*100 /(8*10)) from the 120 patches.

\subsubsection {Representation of Frequency Band Envelop}
We take patches of 8 mel-bins across time to model energy modulations at different frequency bands across time. This captures variations/behavior of energy in a frequency band across time. For a spectrogram, 12 such patches are obtained by equally dividing the frequency axis. These have the input dimension of 8 x 100 and obtain a bottleneck feature $e_F^{fenv}$ as shown in Fig. 1.  We thus obtain a total of 12 input bottleneck features for a given 1s spectrogram input.  

\subsubsection{Representation of Spectral Envelop} Similar to frequency band envelope, we learn embeddings to characteristics of spectral envelope across time. For this we take patches across 100ms encompassing the entire frequency spread of 96 mel-bins, thus having the dimension of 96 x 10. Again these patches are chosen sequentially and we obtain a bottleneck feature $e_F^{env}$. We get a total of 10 bottleneck vectors for any given input spectra.

\subsubsection{Representation of the Whole Spectogram} Finally, given an input mel-spectrogram, we obtain its global characteristics. This is done by down-sampling and resizing the input spectrogram from 96x100 to 8x10, done by using bi-cubic interpolation. This is mainly done to capture at a very coarse level how the spectrogram "looks". We again follow a similar recipe of obtaining a bottleneck feature and call it $e_F^{o}$. We obtain only 1 bottleneck vector for every input spectra. Thus, in total, we obtain 143 latent vectors, (120+12+10+1) that describe the input spectrogram at various levels. 
\subsection{Clustering, Code Book Generation, \& Embeddings}
For each family of the bottleneck features obtained, we do k-means clustering with the number of clusters fixed as D, which is a hyper-parameter. This can be easily implemented using popularly available libraries e.g. sklearn. For any given spectrogram, the goal is to obtain a discrete feature representation that can capture all of the features that we have explained above. Given any input spectra, we first obtain a total of 143 latent codes as described above at a chosen bottleneck factor $F$. These embeddings are clustered using k-means separately using each of the families to which they belong, i.e. we cluster $i^{th}$ entry in each of the embedding families, $e_{F}^{o}(i)$ separately from $e_{F}^{env}(i)$. For each of the learned k-means mapping, we obtain a feature vector $f$ of dimension $D$, similar to a bag-of-words representation i.e. each feature vector contains the counts of how many times the embeddings fall into the bucket of the cluster centroids. Thus we obtain a $D$ dimension vector for each of the four categories mentioned thus obtaining, $f_F^{o}$, $f_F^{env}$, $f_F^{fenv}$ and $f_F^{pat}$. Each of these features $f$ contains counts of how often the code-word of embedding from the audio spectrogram is present in a given 1s input. As expected the sum of the feature vector would be 143 equal to the total number of embeddings we have from the four families of features.  This is now our representative feature vector for input instead of a spectrogram.

\subsection{Classification Setup \& Input Code Masking} We concatenate all of the feature vectors $f_F^{o}$, $f_F^{env}$, $f_F^{fenv}$ and $f_F^{pat}$ to obtain a feature vector $F_v$ for a particular input spectrogram. This is a $4*D$ dimensional input and is used along with the label present in the dataset. As proposed in work by \cite{chen2020simple}, we employ just MLP classification heads. Since we have a multi-label classification problem, use sigmoid as our final activation function, with Huber loss as our error criterion to minimize between the actual and the predicted samples, with Adam optimizer \cite{kingma2014adam}. We improve the robustness further by randomly masking input features by a chosen probability $p$. This is as simple as replacing the actual count of the feature vector by 0 to remove the contribution of a code-book index. This makes our predictions more robust to outliers and as we will see improves the performance of our models. All of the training was carried out using Tensorflow framework \cite{45381}.

\label{sec:method}
\section{RESULTS}
\label{sec:results}
Given a feature vector $F_v$ of dimension equal to 4*dimension of the number of codewords chosen $D$, we have as our input of dimension $F_v = 4*D$. As described in the previous section $F_v$ is the concatenation of individual feature vectors obtained from all of the four features. We deploy the same strategy as what was proposed for classification head given a feature vector $F_v$ in \cite{chen2020simple}. We experiment with several combinations of hyper-parameters namely the size of the feed-forward classification head with the given labels for audio of dimension 200 with 1/0 indicating presence/absence of the category i.e. small namely 2-layer 256 dimensional fully connected architecture and large namely 2-layer 4096 dimensional model. Note that \cite{chen2020simple} used a 2-layer 2048 neuron MLP architecture. This was carried out for all 4 different sizes of code-book $D$: 16, 64, 256, 1024. For all of the proposed models as described above, we tweak the dropout rate \cite{srivastava2014dropout} as a form of regularization with $0.1$ and $0.4$. The results are shown in the Figure 2.
\begin{figure}[!hbt] \centering \includegraphics[width=8.7cm,height=5cm]{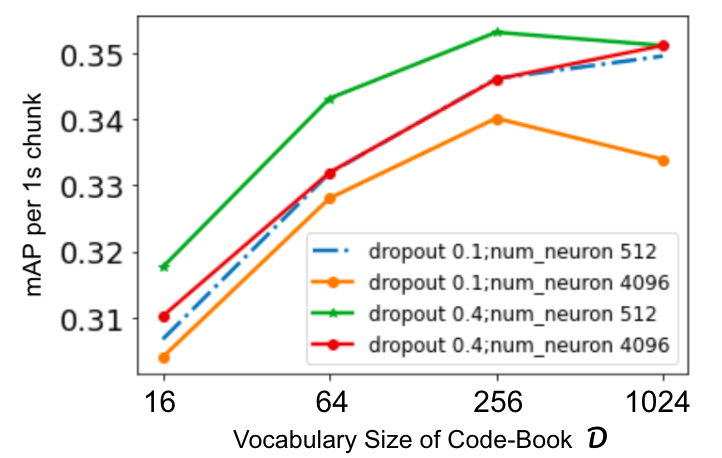} \caption{ mAP on the validation set of FSD 50K. For each dim D code book, MLP head w.r.t dropout rate and the number of neurons in each layer for F =10 is shown. We achieve the best value of 0.35 vs 0.39 of a Transformer architecture in \cite{verma2021audio}. A compression factor of 20 was consistently under-performed the factor 10 and is not shown here for clarity.} \label{fig:slot-fill}\end{figure} 
We see from the plot, the best mAP score is for learned 256 code words, with the classification head having 512 neurons in each layer with a dropout rate of 0.4 and a compression factor of 10. A compression factor of 20 was consistently under-performed the factor 10 and is not shown here for clarity. The choice of an intermediate number of codewords makes sense: Too few of the code-words may not be able to capture all of the variations present in the data, and too large of codewords yields to over-fitting and learning finer unnecessary details. The same holds as expected for the size of classification head, smaller models yielding much better performance as opposed to a large model possibly over-fitting the dataset. With the best performing architecture keeping it fixed, we improve the performance of the models further by making models robust to errors/outliers in various input codes. 

\begin{figure}[!hbt] \centering \includegraphics[width=0.8\linewidth,height=4.5cm]{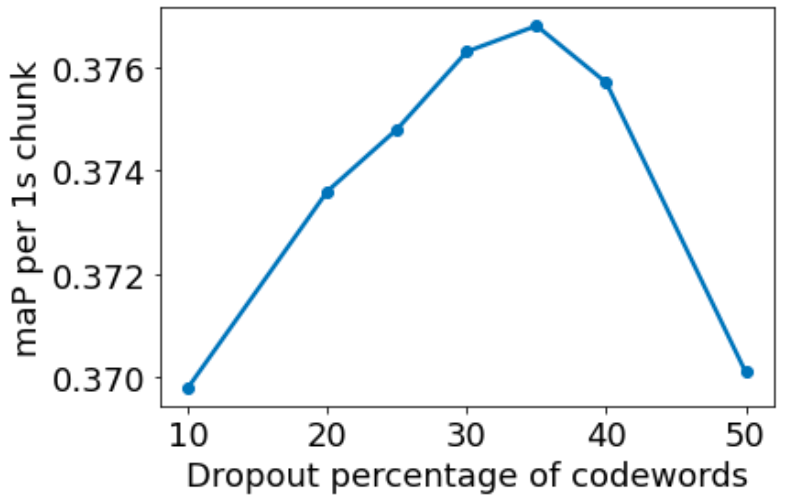} \caption{Effect of probability of dropped input tokens on mAP scores for 1s audio chunks for best performing model. Notice that there exists a optimum percentage to maximize the performance, and it was tuned over 10, 20, 25, 30, 35, 40 and 50 percent of the dropped tokens over the feature vector of size $4*D$. } \label{fig:slot-fill}\end{figure} 

We randomly drop input features by a probability $p$ and see the performance, in terms of mean average precision. Masking of inputs is not new and has been used in various contexts. As shown in BERT \cite{devlin2018bert}, it adds robustness to the learned embeddings. We see that there exists an optimum value of the number of tokens one should drop: too many tokens being dropped and we lose valuable information. Too few tokens dropped and we do not build the model agnostic to outliers. We are perhaps the first ones to introduce this idea in a bag-of-words setup. We obtain the best overall performance of about 0.38 mean average precision (mAP) score with 35\% randomly dropped tokens. We also would like to note that there can exist many more values/parameters that can be tuned.  

Finally, in Table 1 we report clip level scores which are obtained by averaging the probability scores as reported in \cite{fonseca2020fsd50k}. We come strikingly close to baseline Transformer models, and outperform strong state of the art models as recent as 2018! It is noteable as we use all of the tools that could have been used/had access to as early as 2006.  Similar to how the Transformers were improved upon from the baseline model (multi-resolution, pooling, larger models), our current model can also be improved upon in several ways, a few of them have been described in future work. We report the best-performing model with that of convolutional neural architectures and Transformer models. Our setup achieves a comparable performance which is fascinating in itself. However, the goal of this work is simplicity: To showcase how to obtain comparable performance without utilizing any convolutional, transformer, attention, rnns, or mixer blocks \cite{tolstikhin2021mlp} going against the mainstream research as a first step in this line of work, of using embeddings. 
\begin{table}[ht]
  \caption{\itshape Comparison of proposed architecture as shown in the table below for mAP metric at clip level. Our approach can outperform widely used CNNs by significant margins and comes close to the baseline Transformer models. \cite{fonseca2020fsd50k}. Similar to the improvements over baseline Transformer architectures, we can in future also improve the performance in a variety of ways.  }
	\centering
	\begin{tabular}{|c|c|}
		\hline
		Neural Model Architecture & mAP\\\hline
		CRNN \cite{fonseca2020fsd50k} & 0.41\\
		VGG-like \cite{fonseca2020fsd50k} & 0.43 \\
		ResNet-18 \cite{fonseca2020fsd50k} & 0.37 \\
		DenseNet-121 \cite{fonseca2020fsd50k} & 0.42 \\ 
		Baseline/Wavelet Transformer \cite{verma2021audio}  & 0.46/0.54\\\hline
		Current Work & 0.44
        \\\hline
	\end{tabular}
	\label{tab:example}
\end{table}
 
\section{Conclusion and Future Work}
\label{sec: conclusion} We show the power of vanilla embeddings followed by code-book learning to yield representations in clustered space of the signal of interest. We achieve this by modeling the envelopes across different frequency bins, spectral envelops, patches as well as the overall spectrogram at a coarser scale. This idea can be strengthened in the future by deploying more sophisticated clustering algorithms than k-means, such as UMAP \cite{mcinnes2018umap}, DBSCAN \cite{ester1996density} or better end to end learned code-book  such as neural discrete representation learning \cite{oord2017neural}. Due to the sheer number of parameters and limited computing resources, we could not achieve large-scale hyper-parameter tuning beyond what is reported, although it would certainly help improving mAP scores. Further, bringing temporally statistics such as n-grams will help boost the performance too, similar to gains showed over a simple bag of words approach in natural language processing \cite{suzuki2002language}.  

\section*{Acknowledgments}
The author thanks Prof. J. K. Verma for help in proof reading the manuscript, the Department of Mathematics, IIT Bombay for providing facilities. He is thankful for Prof. Stephen Boyd's lectures in Stanford ENGR 108 \cite{engr108} for sowing seeds of this work, and Robert Crown Library at SLS. 

\balance
\bibliographystyle{IEEEbib}
\bibliography{bib}

\end{document}